\newcommand\Mdot{\mathaccent 95 M}
\newcommand\lax{\>\vcenter{\hbox{$<$\hskip-.75em\lower1.0ex\hbox{$\sim$}}}\>}
\newcommand\gax{\>\vcenter{\hbox{$>$\hskip-.75em\lower1.0ex\hbox{$\sim$}}}\>}

 \documentstyle[11pt,paspconf]{article}        

\begin{document}

\title{The Quasi-Coherent Oscillations of SS~Cygni}
\author{Christopher W.\ Mauche}
\affil{Lawrence Livermore National Laboratory,\\
L-41, P.O.\ Box 808, Livermore, CA 94550}

\vbox to -12pt{\vskip -5.50cm
\hbox to \hsize{{\it 1996, IAU Colloquium No.~163---Accretion
                     Phenomena \& Related Outflows,}\hfil }
\hbox to \hsize{{\it ed.\ D.~Wickramasinghe, L.~Ferrario, \&
                     G.~Bicknell (San Francisco: ASP)}\hfil }
\vss}
\vskip -12pt

\begin{abstract}
Properties of the quasi-coherent oscillations in the extreme ultraviolet flux
of the dwarf nova SS~Cygni are described.
\end{abstract}



\section{Introduction}

Rapid periodic oscillations are observed in the optical and soft X-ray flux of
high accretion rate cataclysmic variables (CVs; nova-like variables and dwarf
novae in outburst) (Patterson 1981; Warner 1995a; 1995b). These so-called
``dwarf nova oscillations'' (DNOs) have high coherence ($Q\approx 10^4$--$10^6$),
periods of $\approx 10$--30~s, and amplitudes of $\approx 10$--30\% in soft
X-rays and $\lax 0.5$\% in the optical. DNOs have never been detected in dwarf
novae in quiescence, despite extensive searches; they appear on the rising
branch of the dwarf nova outburst, typically persist through maximum, and
disappear on the declining branch of the outburst. The period of the oscillation
also correlates with outburst state, decreasing on the rising branch and
increasing on the declining branch.

The dwarf nova SS~Cygni routinely exhibits DNOs during outburst. Optical
oscillations have been detected at various times with periods ranging from
7.3 s to 10.9~s (Patterson, Robinson, \& Kiplinger 1978; Horne \& Gomer 1980;
Hildebrand, Spillar, \& Stiening 1981; Patterson 1981). At soft X-ray
energies, oscillations have been detected in {\it HEAO~1\/} LED~1 data at
periods of $\approx 9$~s and 11~s (C\'ordova et~al.\ 1980; 1984) and in {\it
EXOSAT\/} LE data at periods between 7.4~s and 10.4~s (Jones \& Watson 1992).
Here, we describe the properties of the oscillations in the extreme ultraviolet
(EUV) flux detected with the {\it Extreme Ultraviolet Explorer\/} ({\it EUVE\/})
Deep Survey (DS) photometer and Short Wavelength (SW) spectrometer during
target-of-opportunity observations of SS~Cyg in outburst in 1993 August and
1994 June/July. For additional details, see Mauche, Raymond, \& Mattei (1995)
and Mauche (1996).

\section{Relation between Period and Mass-Accretion Rate}

Oscillations in the EUV flux were detected by calculating power spectra of
DS count rate light curves with 1~s time resolution. The period of the
oscillation as a function of time and of the log of the 75--120~\AA \ SW count
rate is shown in Figure~\ref{fig1}.
\begin{figure}
 \vspace{7.5cm}        
\caption{SW count rate ({\it filled circles with error bars\/}) and period
({\it open circles\/}) as a function of time for the ({\it a\/}) 1993 and
({\it b\/}) 1994 outbursts of SS~Cyg. Period as a function of the SW count
rate for the ({\it c\/}) 1993 and ({\it d\/}) 1994 outbursts.}
\label{fig1}
\end{figure}
These figures demonstrate that the period $P$ of the oscillation anticorrelates
with the SW count rate $I_{\rm EUV}$, but an even stronger statement can be
made. Because the EUV spectrum evolves homologously with time (Mauche, Raymond,
\& Mattei 1995), its bolometric correction is constant, and hence we conclude
that the period anticorrelates with the EUV/soft X-ray luminosity, and, by
inference, with the mass-accretion rate $\Mdot $ onto the white dwarf: an
unweighted fit to the combined data gives $P=7.08\,I_{\rm EUV}^{-0.094}~{\rm s}
\propto \Mdot ^{-0.094}$. In contrast, the period of the QPOs of the transient
Be-neutron star binary A0535+262 scale like $P\propto \Mdot ^{-0.46}$ (Finger,
Wilson, \& Harmon 1996), in good agreement with the theory of disk accretion
onto a magnetized star: $P\propto \Mdot ^{-3/7}$ (Ghosh \& Lamb 1991). For such
a model to apply to SS~Cyg, the field must be an effective high-order multipole
($l\approx 7$) with a strength at the surface of the white dwarf of 0.1--1 MG
(Mauche 1996).

\section{Mean Power Spectra and Waveforms}

Unlike the QPOs of A0535+262 and other neutron star binaries, the oscillations
of SS~Cyg are quite coherent. The slowest change of period with time occurred
during the plateau of the 1993 outburst, when the coherence $Q\equiv |\Delta
P/\Delta t|^{-1} = (0.083~{\rm s}/2.70~{\rm d})^{-1} = 3\times 10^6$; the
fastest change of period with time occurred during the rise of the 1994 outburst,
when $Q = (1.7~{\rm s}/0.78~{\rm d})^{-1} = 4\times 10^4$; on orbit-to-orbit
timescales, $Q\gax (0.03~{\rm s}/30~{\rm min})^{-1} = 6\times 10^4$. The
coherence of the oscillation can be seen in the mean power spectra shown in
Figures~\ref{fig2}({\it a\/},{\it b\/})
\begin{figure}
 \vspace{7.5cm}        
\caption{Mean power spectra of the ({\it a\/}) 1993 and ({\it b\/}) 1994
outbursts of SS~Cyg. Mean waveforms ({\it filled circles\/}) and residuals
relative to the fitted sinusoid ({\it crosses\/}) for the ({\it c\/}) 1993
and ({\it d\/}) 1994 outbursts.}
\label{fig2}
\end{figure}
constructed by adding the power spectra of the individual satellite orbits
after scaling by the local oscillation period. In addition, these spectra
demonstrate that the mean waveforms are sinusoids to high degree: very little
power is observed in the first harmonic, and no power is observed in the second
or subharmonics. Indeed, the additional power in the second harmonic of the
power spectrum of the 1994 outburst is due to the interval near the peak of
the outburst ($P< 7.5$~s); during the plateau phase of that outburst ($\rm JD >
2449532$), the power spectrum is actually cleaner than the power spectrum of the
1993 outburst. The peak power in the first harmonic relative to the fundamental
is 2.5\% for the 1993 outburst, 2.7\% for the 1994 outburst, 9.0\% for the peak
of the 1994 outburst, and 1.8\% for the plateau of the 1994 outburst. The mean
waveforms of the 1993 and 1994 outbursts are shown in
Figures~\ref{fig2}({\it c\/},{\it d\/}). These figures demonstrate that the mean
waveforms are surprisingly stable, and can be fit well by a function of the form
$A + B\sin2\pi\phi - C\cos4\pi\phi$. For the 1993 outburst, $B/A=17.1\%$ and
$C/B=12.3\%$; for the 1994 outburst, $B/A=16.5\%$ and $C/B=14.1\%$; for the peak
of the 1994 outburst, $B/A=11.4\%$ and $C/B=27.9\%$; for the plateau of the 1994
outburst, $B/A=19.1\%$ and $C/B=11.0\%$. Evidently, during the peak of the 1994
outburst, the amplitude of the oscillation is reduced and the waveform is
distorted, with more flux coming out in the first harmonic. 

\section{Spin Period of the White Dwarf?}

Careful examination of the power spectrum of the 1994 outburst reveals a narrow
peak at $\nu/\nu_0\approx 0.09$ which is also present in the direct mean power
spectrum at $\nu=0.012\pm 0.002$~Hz (FWZI). This feature is present above the
noise in some orbits and not in others, and its period is relatively but not
strictly constant. Because this feature is so far displaced from the
fundamental, and because it persists in power spectra derived from light curves
with 10~s bins, it is unlikely that this oscillation is an artifact of the
dwarf nova oscillation. Perhaps it is the spin frequency of the white dwarf.
If so, the corotation radius $r_{\rm co}\equiv [GM_* /(2\pi\nu_* )^2]^{1/3}
\approx 3\times 10^9~{\rm cm}\sim 8$ white dwarf radii. In contrast, if the
magneto\-spheric model applies to SS~Cyg, the inner edge of the disk is at
$r_0 = 5.6\times 10^8~{\rm cm}\approx 1.5$ white dwarf radii at the peak of the
outburst. The resulting fastness parameter $\omega _{\rm s}\equiv \nu_* /\nu
(r_0) \approx 0.09$. This value is significantly less than the equilibrium
fastness parameter $\omega _{\rm c}=0.7$--0.95 (Li \& Wang 1996), but it is
unlikely that the white dwarf is in spin equilibrium during outburst---it is
much more likely that it is being spun up. The implied rotation velocity of the
white dwarf is $v=2\pi R_* \nu_* \approx 300~\rm km~s^{-1}$. This compares with
$v\, \sin i < 200~\rm km~s^{-1}$ for U~Gem and $\approx 600~\rm km~s^{-1}$ for
VW~Hyi (Sion et~al.\ 1994; 1995). These values are more meaningfully expressed
as a fraction of the breakup velocity of the white dwarf, $v_{\rm break} = (GM_*
/R_* )^{1/2}$: $v/v_{\rm break} < 4\%$ for U~Gem, $\approx 5\%$ for SS~Cyg, and
$\approx 20\%$ for VW~Hyi. The corresponding ratio of the boundary layer to
accretion disk luminosity is $\zeta = [1-(v/v_{\rm break})]^2 > 0.92$ for U~Gem,
$\approx 0.90$ for SS~Cyg, and $\approx 0.64$ for VW~Hyi. The measured values of
these ratios are $\zeta\approx 0.45$ for U~Gem (Long et~al.\ 1996), $\lax 0.07$
for SS~Cyg (Mauche, Raymond, \& Mattei 1995), and $\sim 0.04$ for VW~Hyi (Mauche
et~al.\ 1991). Apparently, some other mechanism beyond rotation of the white
dwarf is responsible for the low boundary layer luminosities of non-magnetic
CVs.

\acknowledgments

This work was performed under the auspices of the U.S.~Department of Energy
by Lawrence Livermore National Laboratory under contract No.~W-7405-Eng-48. 


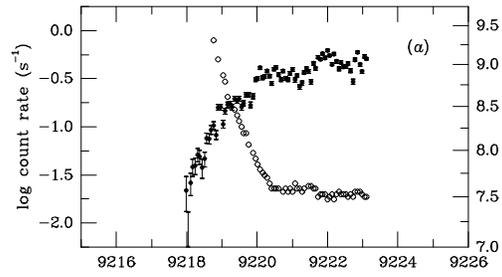
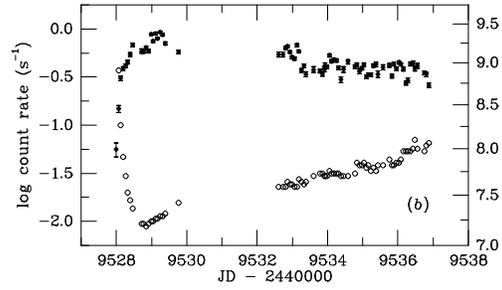
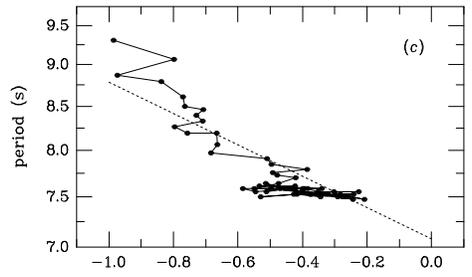
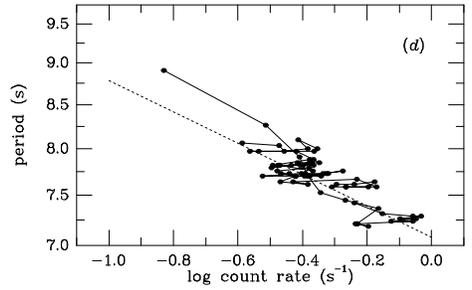
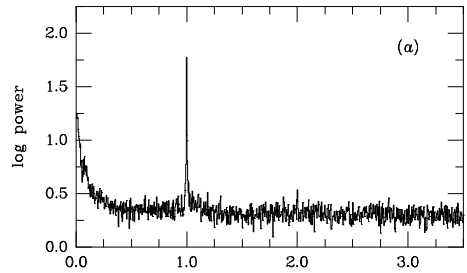
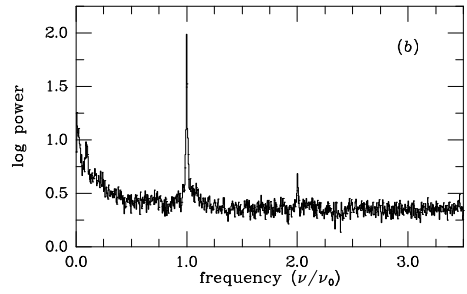
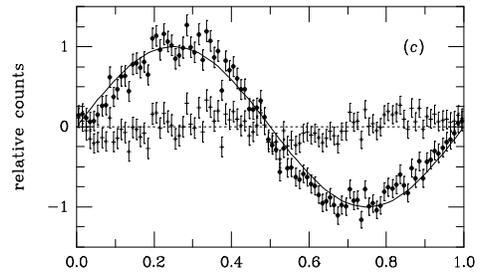
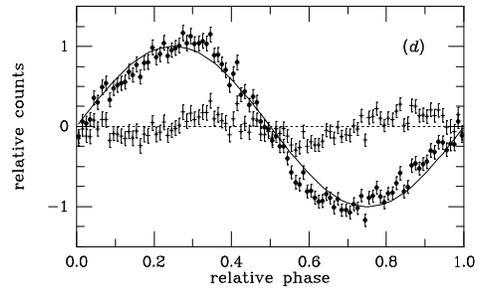
\end{document}